\documentclass[twocolumn,preprintnumbers,amsmath,amssymb]{revtex4-1}
\usepackage{graphicx}
\usepackage{dcolumn}
\usepackage{bm}
\usepackage{color}

\begin{document}
\title{Phase nucleation in curved space}

\author{Leopoldo R. G\'omez$^{1}$}
\email{lgomez@uns.edu.ar}
\author{Nicol\'as A. Garc\'ia$^{1}$}
\author{Vincenzo Vitelli$^{2}$}
\author{Jos\'e Lorenzana$^{3}$}
\author{Daniel A. Vega$^{1}$}

\affiliation{$^1$ Department of Physics, Universidad Nacional del
Sur - IFISUR - CONICET, 8000 Bah\'ia Blanca, Argentina. $^2$ Instituut-Lorentz, Universiteit Leiden, 2300 RA
Leiden, The Netherlands. $^3$ Institute for
Complex Systems Consiglio Nazionale delle Ricerche, and Physics
Department, University of Rome La Sapienza, I-00185 Rome, Italy.}

\date{\today}

\begin{abstract}
Nucleation and growth is the dominant relaxation mechanism
driving first order phase transitions. In two-dimensional flat systems nucleation has
been applied to a wide range of problems in physics, chemistry
and biology. Here we study nucleation and growth of two-dimensional phases lying on
curved surfaces and show that
curvature modify both, critical sizes of nuclei and paths
towards the equilibrium phase. In curved space nucleation and growth becomes
inherently inhomogeneous and critical nuclei form faster on
regions of positive Gaussian curvature. Substrates of varying shape
display complex energy landscapes with several
geometry-induced local minima, where initially propagating nuclei
become stabilized and trapped by the underlying curvature.
\end{abstract}

\maketitle

\section*{Introduction}

In its classical picture, nucleation and growth  (NG)
starts with local fluctuations of an initial metastable phase leading to the
formation of a small nucleus of the equilibrium phase
\cite{Kashchiev}-\cite{Tosatti}. Such
nucleus involves local changes in the free energy of the system. On one hand,
the formation of a nucleus of the less energetic phase produces a decrease in the total free energy scaling with its volume.
On the other hand, it raises the energy by an amount that scales with the area of the interface separating the two phases.

In the case of flat two dimensional systems, the variation in the
free energy due to the formation of a nucleus of radius $R$ takes
the expression \cite{Kashchiev}-\cite{Tosatti}:
\begin{equation}
\Delta F= 2 \pi R \sigma - \pi R^2 \Delta f,
\end{equation}
where $\sigma$ represents the surface tension and $\Delta f$  the
difference in the local free energies of initial and final phases
driving the phase transition. The competition of surface and
volume terms produces the activated dynamics of NG. Only
those fluctuating nuclei whose size overpass the critical value
$R_\textrm{c}=\sigma/\Delta f$ will propagate. All
other nuclei collapse due to surface energy.


On the other hand, phases
lying on curved backgrounds \cite{Nelson}-\cite{BowickAdvPhys} are
not only commonly found in nature in systems like viral capsids,
pollen grains, radiolaria, and others, but they can also be obtained in
the laboratory in the form of soft crystal or liquid crystal phases by
using colloidal particles \cite{Bausch}-\cite{IrvineNatMat}, block
copolymers \cite{HexemerThesis}-\cite{VegaSoftmatter}, liquid
crystals \cite{FernandezNievesNatPhys}-\cite{FernandezNievesPNAS},
and other self-assembled systems. Examples of phase nucleation on
curved geometries in hard-condensed matter may include, among others, graphene \cite{Naumovets}, \cite{LeRoy},
epitaxial growth of helium \cite{Balibar}, and Wigner crystals on rough surfaces \cite{Dahm}.

In equilibrium, these phases display regular structures strongly
coupled to the underlying geometry that are not only ideal model systems to
study the interplay between order and curvature, but also potential platforms for technological applications such as soft litography
\cite{HexemerThesis}, \cite{VegaSoftmatter} or defect
functionalization \cite{DeVries}. Although experiments
and theoretical calculations have contributed to unveil
equilibrium configurations and energetics of topological defects in curved space,
dynamical processes like crystallization and melting still remain
marginally explored \cite{GomezSmectics}-\cite{Meng}.

In this work we obtain generic features of NG in two-dimensional systems lying on arbitrary curved
backgrounds. Our model is based on a Ginzburg-Landau free energy functional which is sufficiently general that can be applied to various physical systems and substrate's geometries. For nearly-flat surfaces and geometries of constant curvature we obtain all the relevant information related to NG in a closed analytical form. On more complex surfaces of varying curvature we explore the effects of the underlying geometry on NG by using simulations. We show that in general the free energy barrier needed to form a critical nuclei is strongly dependent on the underlying curvature, being smaller (larger) for region of positive (negative) curvature. As a consequence, in geometries of varying curvature NG is an inhomogeneous process, first starting in regions of positive curvature.
Also, due to the competing effects of positive and negative curvature, geometries of varying curvature are shown to display complex relaxation free energy landscapes, with several barriers and local minima, where initially propagating nuclei become stabilized and trapped by the underlying curvature. Additional effects coming from system's thickness and strain contributions during crystallization are also considered and discussed in view of recent experimental work.

\section*{Results}

\textbf{Model.} Here we present a model to study how the underlying geometry
affects the laws of NG in curved space. In general, the geometry of an oriented surface (i.e. separating an ``internal
region" from an ``external region" of three dimensional space) can be completely characterized by
the principal curvature maps $\kappa_i({\bf r})=\pm 1 /R_i$ ($i=1,2$),
where $R_1$ ($R_2$) is the radius of the smallest (largest) circle
tangent to the surface at ${\bf r}$, and the sign is positive
(negative) if the circle is contained in the internal (external)
region. Alternatively one can use the Gaussian
$K({\bf r}) = \kappa_1({\bf r}) \kappa_2({\bf r})$, and mean $H({\bf r})=(\kappa_1({\bf r})+\kappa_2({\bf r}))/2$ curvature
\cite{Struik}.

The case $K=0$ and $H\ne 0$ represents a plane bent in only one
direction. Since this surface can be flattened preserving the areas (Minding's theorem \cite{Struik}),
nucleation can be mapped to the case of a planar surface and thus results
to be trivial. Non-trivial cases have non-zero Gaussian curvature but
may have zero mean curvature as in the case $\kappa_1=-\kappa_2$ which
represents a saddle. This clearly can not be flattened preserving
the areas and thus nucleation has to be revisited.  In other words, the Gaussian represents an ``intrinsic''
curvature and, as we shall see,  results to be most important in describing
physical processes on the surface. Below we show how
the Gaussian curvature influences NG in curved geometries.

The mean curvature is not only related with the
geometry of the surface, but it is also tied to the definition of
``internal'' and ``external'' regions.  For example, while crest have
positive mean curvature and valleys have negative mean curvature these
values can be inverted by exchanging the definition of ``internal'' and
``external'' regions \cite{Struik}.
As we will see later, mean curvature will
modify NG for 2D systems with thickness.

We study the dynamics of first order phase transitions through a
scalar and real order parameter $\psi(\textbf{r},t)$. Points on the
surface are specified by a system of curvilinear coordinates
$\textbf{r}=(x_1,x_2)$. In these coordinates an infinitesimal arc length $ds$
is given in Einstein notation by $ds^2\equiv|d\textbf{r}|^2=g_{\alpha\beta}dx^\alpha dx^\beta
$, where $g_{\alpha\beta}$ is the metric tensor. Depending on
the physical system considered, the order parameter is or is not a
conserved quantity \cite{ChaikinLubensky}. For example, while conserved order parameters
are required to study nucleation of binary mixtures or solid
solutions, non-conserved order parameters are commonly used in
studies of the liquid to crystal transition, or in magnetic and
spin-related phases.

In general, the free energy of a mixed state in a system lying on
a curved surface can be expanded in terms of the order parameter:
\begin{equation}
\label{freeenergyfunctional} \frac{F}{k_\textrm{B} T}=\int  d^2r \, \sqrt{g} \,
[f(\psi)+\frac{D}{2} \, g^{\alpha \beta} \, \partial_\alpha
\psi(\textbf{r}) \,
\partial_\beta \, \psi(\textbf{r})],
\end{equation}
where $k_\textrm{B}$ is the Boltzman's constant, $T$ the temperature and $g$
 is the determinant of the metric tensor \cite{Struik}.

In a curved surface the differential of area is $d^2r\sqrt{g}\equiv
dx_1dx_2\sqrt{g} $ so the first term in the free energy represents the
contribution from a local homogeneous situation, where $f(\psi)$ is
the local free energy areal density of a phase with an order parameter
$\psi$. For a two-phase
system this term takes the typical double well form
\cite{Chan}-\cite{Imawatsu}, with two local minima corresponding
to the initial ($\psi=0$) and final ($\psi=1$) phases, separated by a local free energy barrier. This can be written phenomenologically as:
\begin{equation}
f(\psi)=\frac{1}{4} \, \eta \psi^2 (\psi-1)^2+\frac{3
\varepsilon}{2} (\frac{\psi^3}{3}-\frac{\psi^2}{2}),
\label{density}
\end{equation}
where $\eta$ and $\varepsilon$ are constants. Here
the local free-energy difference driving the phase transition,
typically controlled by temperature, is given by $\Delta
f=f(0)-f(1)$. Note that $\Delta f\equiv \varepsilon/4$ is proportional to the degree of supercooling in the system.
The second term of Eqn. \ref{freeenergyfunctional} is the curved space generalization of the gradient square term ($|\nabla
\psi|^2$) that penalizes the formation of interfaces \cite{Kashchiev}, \cite{Kelton}.

In this Ginzburg-Landau approach the dynamics of the phase
transition can be studied through a relaxational equation of the
form:
\begin{equation} \tau \frac{\partial \psi}{\partial
t}=-(i \,  \triangle_\textrm{LB})^{2n}\{\frac{\delta F}{\delta \psi}\},
\end{equation}
where $\tau$ is a characteristic time scale (for simplicity here
we take $\tau \equiv 1$), $i=\sqrt{-1}$, $n \equiv 0,1$ for
non-conserved and conserved order parameters respectively, and
$\Delta_\textrm{LB}= \frac{1}{\sqrt{g}}\frac{\partial}{\partial
x^i}(g^{ij}\,\sqrt{g} \, \frac{\partial}{\partial x^j})$ is the
Laplace-Beltrami operator, which is the curved-space
generalization of the Laplacian \cite{GomezSmectics},
\cite{Garcia}.

Note that a more complex evolution
equation in needed for conserved order parameters. For
simplicity we concentrate on the time evolution of non-conserved
order parameters. However, most results related to NG on curved geometries can be applied to both \cite{ChaikinLubensky}, \cite{cahn}.

\begin{center}
\begin{figure}[t]
\includegraphics[width=6.5 cm]{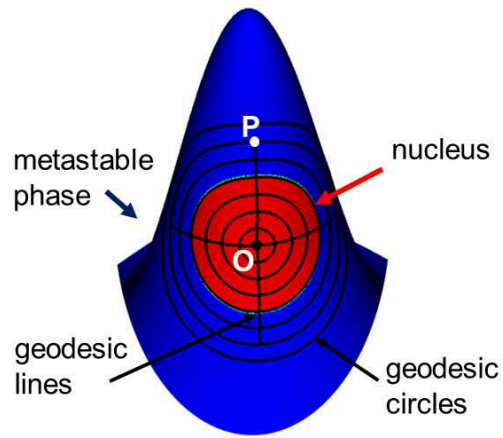}
\caption{\textbf{Geodesic polar coordinates.} In this work geodesic polar coordinates are used to describe the
growth or collapse of a nucleus (red domain) in the sea of the
metastable initial phase (blue). Here a point $P$ on the surface
has coordinates $(r,\theta)$, being $r$ the geodesic distance to
the origin $O$ and $\theta$ the angle.}
\end{figure}
\end{center}

\textbf{Critical Nuclei and Growth and Dissolution Laws.} We start by analyzing the fate of a nucleus resulting from a spontaneous fluctuation,
whose time evolution follows Eqn. 4.

In the following we make use of geodesic polar
coordinates, which are the curved-space generalization of the
polar coordinates \cite{Struik}. We start by locating an origin
$O$ in the center of the nucleus (Fig. 1). Then, a point $P$
located on the surface is associated to coordinates $(r,\theta)$.
The coordinate $r$ is defined as the geodesic distance between $P$
and the origin $O$. Closed lines of constant geodesic distance
$r_0$ are called geodesic circles, and at any point are orthogonal
to the geodesics starting at $O$. The coordinate $\theta$ of the
point $P$ is the angle that the geodesic connecting $O$ and $P$
makes with a reference geodesic starting in $O$.

In these coordinates the metric of any surface always take the simple
form $ds^2=dr^2+G(r,\theta)d\theta^2$, where the function
$G(r,\theta)$ depends on the geometry of the surface through the
equation $\frac{\partial^2 \sqrt{G}}{\partial r^2}+K \sqrt{G}=0$,
with $K$ the Gaussian curvature of the surface, and $G(0,\theta)=0$ \cite{Struik}.

When using geodesic polar coordinates the evolution equation 4 can be largely simplified to study steady state solutions representing
the growth or collapse of nuclei during NG. Here, the evolution equation of a steadily propagating nucleus is
decomposed (see methods) in two equations for the interface profile
(Eqn. 5) and the propagation rate (Eqn. 6).
\begin{eqnarray}
 D\frac{d^2 \psi}{dX^2}+v \frac{d\psi}{dX}&=&\frac{\partial
f(\psi)}{\partial \psi}\\
\frac{dR}{dt}+D\kappa_\textrm{g}(R)&=& v
\end{eqnarray}
where the comoving coordinate $X$ is given by $X=r-R(t)$, $R(t)$ is the radius of the nucleus
at time $t$, $v\equiv \sqrt{2\eta/D}3\varepsilon$ a propagation constant, and the geodesic curvature $\kappa_\textrm{g}(r,\theta)=\frac{\partial }{\partial r}
ln\sqrt{G(r,\theta)}$ is the local curvature of the geodesic
circles providing information about the underlying geometry.

It is interesting to note that for any geometry, the equation for the interface
profile is always the same when written in geodesic polar
coordinates. Its solution is exactly the same as found by Chan for
NG on a plane \cite{Chan}-\cite{Imawatsu}:
$\psi(X)=1/(1+exp(1/2\sqrt{2\eta/D}X)$.

By contrast, Eqn. 6 shows that the evolution of nuclei's sizes
is influenced by the underlying geometry through the geodesic
curvature $\kappa_\textrm{g}$. Note that for a planar geometry
$\kappa_\textrm{g}(R)\equiv 1/R$ gives the classical growth rate
for NG on a plane. Although in the general case this is a
complicated non-linear ordinary differential equation, it can be solved in
closed form for simple geometries, as shown below.

In general, a nucleus will grow only if its initial radius
$R_0$ exceeds the critical value $R_\textrm{c}$, obtained by setting
$\frac{dR}{dt}\equiv0$, i. e.:
\begin{equation}
\kappa_{g}(R_\textrm{c})=1/R_\textrm{c}^\textrm{0},
\end{equation}
where $R_\textrm{c}^\textrm{0}=D/v$ represents the critical size for NG on
a planar substrate, under the same external conditions.

Thus, Eqn. 7 shows that on a curved substrate the critical radius
will in general depend on both curvature (through
$\kappa_\textrm{g}$) and supercooling
(through $R_\textrm{c}^\textrm{0}$). A characteristic
length associated to the curvature can be defined by $R_\textrm{K} \sim
1/\sqrt{|K|}$. We can expect different regimes of NG depending on
how large the ratio $R_\textrm{c}^\textrm{0}/R_\textrm{K}$ is.

In the following sections we use Eqns. 4, 6 and 7 to analyze how
the geometry modifies NG on substrates with different
distributions of curvature.

\textbf{Nucleation on nearly flat substrates.} For cases when the critical nucleus is very small as compared with
curvature ($R_\textrm{c}^\textrm{0}/R_\textrm{K}<<1$), equally valid for NG far from the
coexistence line or for slightly curved geometries, the geodesic
curvature can be approximated through a series expansion of the
form $\kappa_\textrm{g}(r) = \frac{1}{r}-\frac{1}{3} K_\textrm{0} r+\ldots$, where
$K_\textrm{0}$ is the Gaussian curvature of the substrate evaluated at the
center of the nucleus.

Using this expansion in combination with the equation for the
critical nuclei (Eqn. 7), we get a critical radius which can be written in the
form:
\begin{equation}
R_\textrm{c}=R_\textrm{c}^\textrm{0} \Gamma_\textrm{K} +\ldots
\end{equation}
where $\Gamma_\textrm{K}$ is a geometric factor given by:
\begin{equation}
\Gamma_\textrm{K}=1-\frac{1}{3} \, K_\textrm{0} \, {R_\textrm{c}^\textrm{0}}^2
\end{equation}
Thus, under same external conditions, the curvature modifies the process of NG and the critical nucleus will be
smaller for regions of positive curvature, like crests and valleys ($K_\textrm{0} > 0$), and larger for negative curved regions like saddles ($K_\textrm{0} < 0$).

The energy needed to form a small nucleus of radius $r$ on the
curved substrate can be written as $\Delta F(r)=- A(r) \, \Delta
f+ P(r) \, \sigma$, where $A(r)\sim \pi r^2-\frac{\pi}{12} K_\textrm{0} r^4
+\ldots$ and $P(r)\sim 2 \pi r - \frac{\pi}{3} K_\textrm{0} r^3 +\ldots$
represent the area and perimeter of a geodesic circle with a first order
correction due to curvature. The height of the free energy barrier
for nucleation $\Delta F_\textrm{c}$ is obtained by evaluating
$\Delta F(r)$ at the critical size, and can also be written as:
\begin{equation}
\Delta F_\textrm{c}=\Delta F_\textrm{c}^\textrm{0} \Gamma_\textrm{K} +\ldots
\end{equation}
where $\Delta F_\textrm{c}^\textrm{0}=2 \pi \sigma R_\textrm{c}^\textrm{0} -\pi \Delta f \, {R_\textrm{c}^\textrm{0}}^2$
is the energy barrier on a planar geometry.

This last expression shows how (locally) the height of the barrier
for nucleation is modified by the underlying geometry, decreasing
(increasing) for regions of positive (negative) Gaussian
curvature. The reason for this is simply geometrical. Given a
small nucleus, the ratio between the perimeter and area enclosed
is smaller for surfaces with positive curvature (and larger for negative
curvature).  Thus, for positive curvature the
contribution of the bulk term of the free energy is larger than the interfacial
term, producing a smaller free energy barrier
for nucleation and a smaller critical size.

Note that the last expression for the free energy barrier for NG in a curved geometry
resembles the classical result of heterogeneous nucleation and growth $\Delta F_\textrm{c}=\Delta F_\textrm{c}^\textrm{0} f(\theta)$, where the free energy barrier $\Delta F_\textrm{c}^\textrm{0}$ is modified  by a function $f(\theta)$ of the wetting contact angle $\theta$ of the nuclei with the underlying substrate \cite{Kashchiev}-\cite{Kelton}. However, in classical heterogeneous nucleation the critical size is not modified by the wetting angle $\theta$. This is different from nucleation on curved space, where the underlying curvature affects both free energy barriers and critical sizes.

These results simply show that the underlying curvature can break the
homogeneity of space, such that NG on a substrate of varying curvature becomes
inherently inhomogeneous. In the following sections we consider NG for more general
cases where the effects of the underlying geometry are much
stronger ($R_\textrm{c}^\textrm{0} /R_\textrm{K} \sim 1$).

\begin{center}
\begin{figure}[t]
\includegraphics[width=8.5 cm]{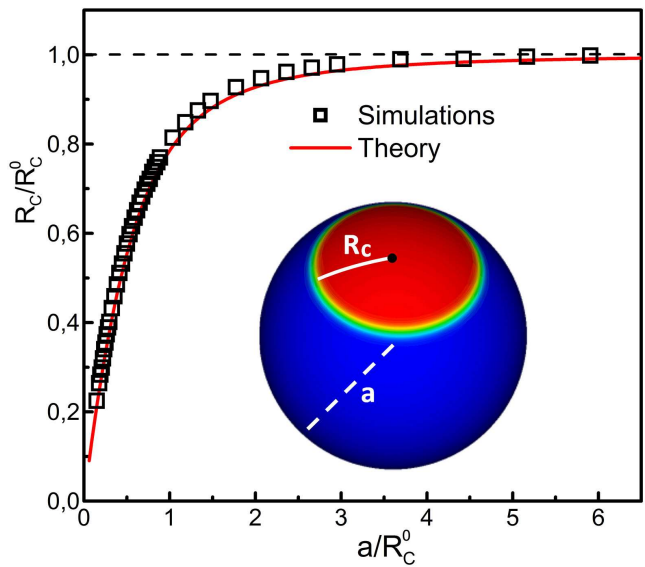}
\caption{\textbf{Critical size for NG on spheres.} The plot shows the critical
(geodesic) radius $R_\textrm{c}$ as a function of sphere radius $a$, both quantities normalized with the critical size for nucleation on a plane $R_\textrm{c}^\textrm{0}$, for simulations
(symbols) and theoretical prediction (line). Note how the critical size decreases when increasing the curvature ($a\longrightarrow 0$). As the underlying curvature decreases ($a\longrightarrow \infty$) the critical size for NG converges to the value find for the flat plane $R_\textrm{c}^\textrm{0}$.}
\end{figure}
\end{center}

\textbf{Nucleation on surfaces of constant curvature.} For spherical substrates the Gaussian curvature is
constant and positive $K=1/a^2$, with $a$ the radius of the
sphere. Here the geodesic curvature is simply written as
$\kappa_\textrm{g}(r)=1/[a \tan(r/a)]$, and the critical size for NG (Eqn. 7) takes
the form:
\begin{equation}
R_\textrm{c}=a \, \arctan(R_\textrm{c}^\textrm{0}/a)
\end{equation}
Thus, the critical size monotonously decreases when decreasing the
radius of the spherical substrate, meaning that for positive
curvature NG is favored for higher curvatures. The reason for this is again geometrical: on the surface of spheres circles have more enclosed area for a given perimeter (as compared to the plane), and the ratio surface-perimeter increases when increasing the curvature.

Given an initial nucleus of size $R_\textrm{0}$, its temporal evolution is
obtained through Eqn. 6, which can be integrated in a closed form giving the implicit
relation $R=R(t)$:
\begin{eqnarray}
vt&=&a \, \sin(R_\textrm{c}/a) \, \cos(R_\textrm{c}/a)
\ln[\frac{\sin((R-R_\textrm{c})/a)}{\sin((R_\textrm{0}-R_\textrm{c})/a)}]\nonumber\\
&&+\cos(R_\textrm{c}/a)^2 (R-R_\textrm{0})
\end{eqnarray}
This is the growth ($R_\textrm{0}>R_\textrm{c}$) or dissolution law ($R_\textrm{0}<R_\textrm{c}$) for
NG on spheres, and reduces to the classical result obtained by
Chan for NG on a flat plane \cite{Chan}-\cite{Imawatsu}, when
taking the Euclidean limit $R_\textrm{c} << a$.

\begin{center}
\begin{figure}[t]
\includegraphics[width=8.35 cm]{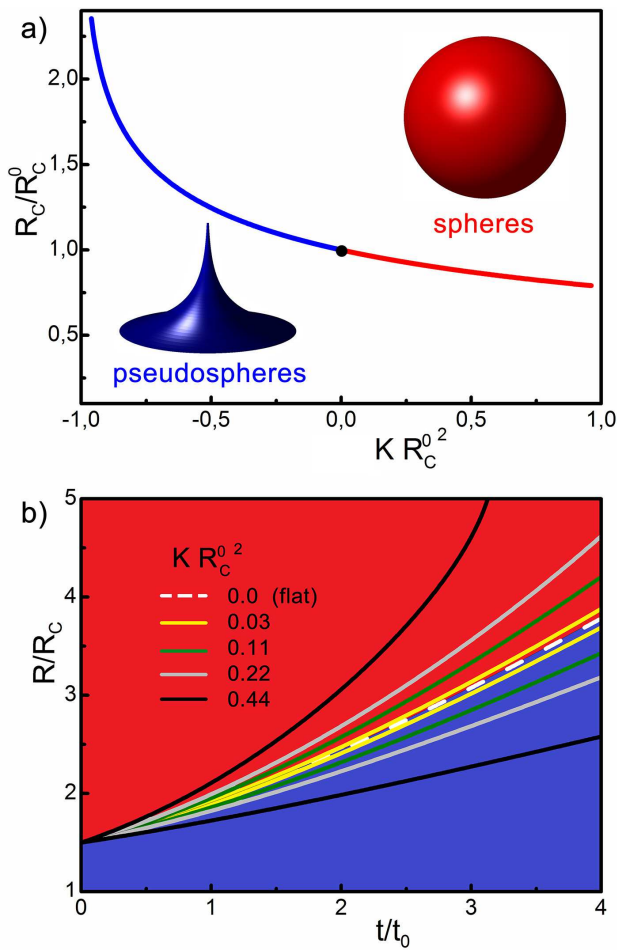}
\caption{\textbf{NG in surfaces of constant curvature.} a) Critical size for nucleation $R_\textrm{c}$ as a function of Gaussian curvature
$K$ for substrates of constant Gaussian curvature, normalized with the critical size on a plane $R_\textrm{c}^\textrm{0}$. The plot shows that the
critical size for NG is larger for pseudospheres ($K<0$, blue line) as
compared with spheres ($K>0$, red line) of the same curvature. Planar substrates ($K=0$),
indicated with the black dot, have critical sizes between spheres
and pseudospheres. b) Time evolution of growing nuclei $R(t)$ on substrates of positive (curves in the red region) and negative (curves in the blue region) constant curvature. Here $R$ and $t$ are normalized with the critical size $R_\textrm{c}$ and $t_\textrm{0}=R_\textrm{c}/v$. Lines in the red (blue) zone represents the growth of nuclei in spheres (pseudospheres).  Continuous lines of the same color represent substrates on the same curvature (in absolute value), and the dashed line represents the growth law on a planar geometry. The growth of nuclei is faster (slower) for substrates of positive (negative) curvature. Note how the evolution converges to the growth law of NG on a plane as the curvature of spheres or pseudospheres decreases.}
\end{figure}
\end{center}

To test the accuracy of these analytical predictions, here we developed
numerical simulations of NG on spherical substrates (see
Methods). Figure 2 shows the excellent comparison between the
theoretical prediction Eqn. 11 (line) and simulations (symbols) for the
critical size $R_\textrm{c}$ as a function of the radius $a$ of the
spherical substrate, normalized with the critical size on a plane $R_\textrm{c}^\textrm{0}$.
This plot clearly shows that nucleation is enhanced for high curvatures (smaller spheres). Note also that as the size of the sphere increases, the critical size for NG converges to the result obtained for the flat plane (vanishing curvature).

We can also consider NG on substrates with constant negative
curvature $K=-1/a^2$, surfaces known as pseudospheres \cite{Struik}, which at
any point display the same local saddle structure. In this case
the geodesic curvature is simply written as $\kappa_\textrm{g}(r)=1/a
\tanh(r/a)$, and the expression for the critical radius takes the
form:
\begin{equation}
R_\textrm{c}=a \tanh^{-1}(R_\textrm{c}^\textrm{0}/a)
\end{equation}
Note that in general all the expression for NG on pseudospheres
are identical to the results obtained for spheres, when replacing
the harmonic by hyperbolic functions. Contrary to spheres, here
the critical size for NG monotonously increase when
decreasing the radius $a$ of the pseudosphere.

Figure 3a shows a comparison for the critical size for nucleation
on spheres and pseudospheres, where the critical radius $R_\textrm{c}$ is
written as a function of the Gaussian curvature $K$ of the
substrate $R_\textrm{c}=1/\sqrt{K}\,\arctan(R_\textrm{c}^\textrm{0}\,\sqrt{K})$ (this expression
is valid for both, positive and negative curvature).
This plot represents
a universal law of NG on substrates of constant curvature, generalizing the well
established results for NG on a plane. Here it is clear that the
critical size for nucleation is bigger for pseudospheres as
compared with spheres of the same curvature. Planar substrates,
indicated with a black dot, have critical sizes between spheres
and pseudospheres.
Note also that Fig. 3a shows that the critical size for nucleation $R_\textrm{c}$ deviates from the result on a planar geometry $R_\textrm{c}^\textrm{0}$ for curvatures of the order $R_\textrm{c}^\textrm{0}/R_\textrm{K} \sim 1$, where for this substrates $R_\textrm{K}\equiv a$, as expected.

Figure 3b shows the time evolution of the size $R(t)$ of growing nuclei on different substrates of positive (curves in the red region), null (dashed curve), and negative (curves in the blue region) constant curvature. This plot clearly shows that the growth of nuclei is faster (slower) for substrates of positive (negative) curvature, as compared to the plane. From the plot it is also clear how the evolution law converges to the law of NG on a plane, as curvature of spheres or pseudospheres decreases (dashed line).

In what follows we consider NG on more complex geometries, where
the varying curvature will produce richer
free energy landscapes and intricate relaxation paths towards
equilibrium.

\textbf{Nucleation on surfaces of varying curvature.} In general, for substrates of non-constant curvature $K$ it can be
difficult to obtain closed analytical expressions for the polar coordinates, needed to describe NG. In addition, for complicated
geometries we can expect the geodesic curvature to be not only function of
the geodesic radius $r$, but also of the polar angle $\theta$,
$\kappa_\textrm{g}=\kappa_\textrm{g}(r,\theta)$. In such cases the breaking of the
radial symmetry makes the simplification of the evolution equation
Eq. 4 much harder, and in general, other parameters related to shape
in addition to size would be needed to describe the nuclei's
evolution.

Given that the critical radius is strongly dependent on
curvature (Fig. 3a), complicated geometries may induce complexities in the
free energy landscape for NG. Also, for substrates of varying curvature
we can expect that the opposite effects of positive and negative
curvature will induce the acceleration or deceleration of
interphases travelling through different regions (Fig. 3b), adding complexity to the
relaxation paths towards the
equilibrium phase.

\begin{center}
\begin{figure*}[t]
\includegraphics[width=15 cm]{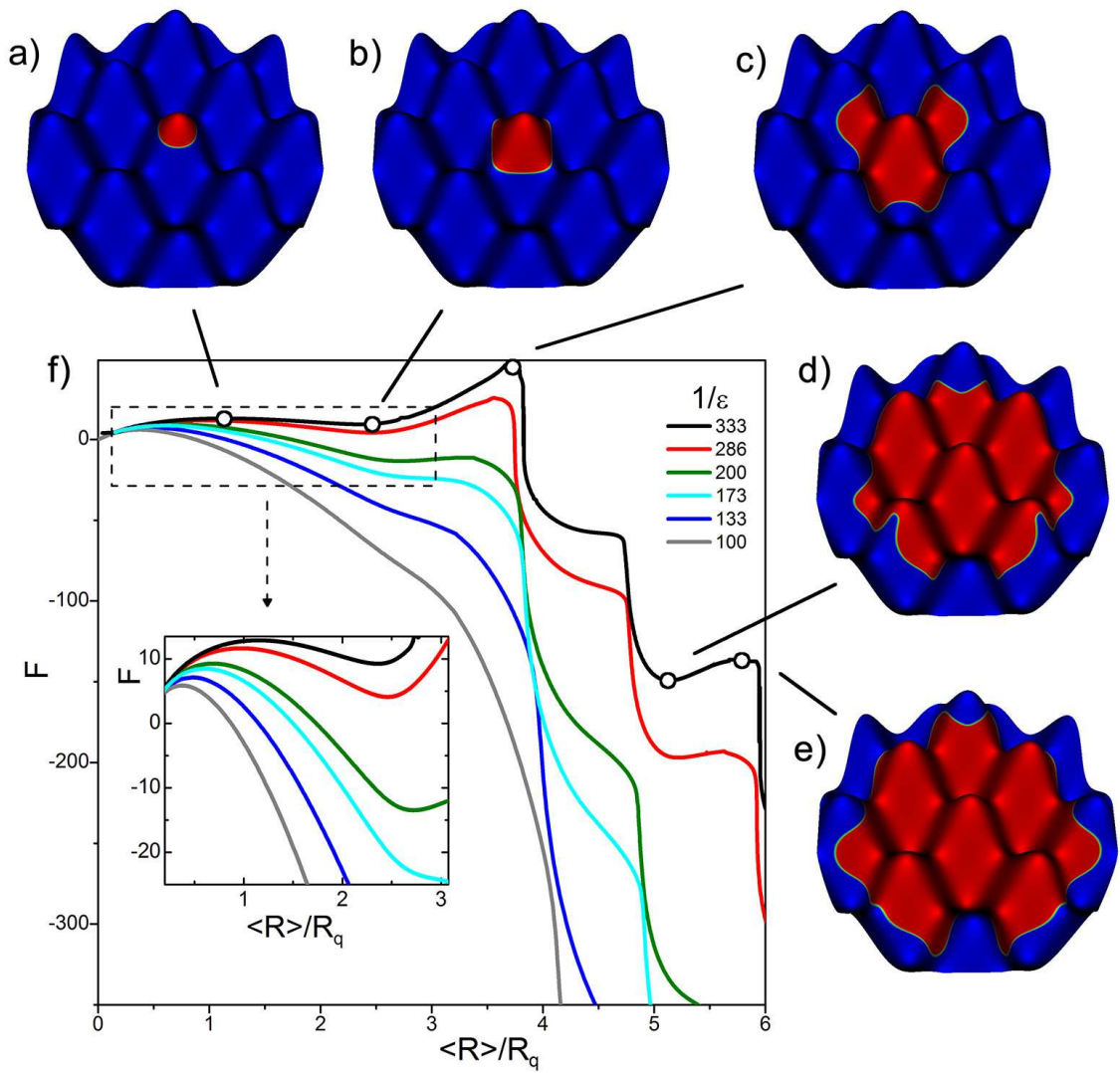}\caption{\textbf{NG
in a sinusoidal geometry.} a), b), c), d) and e) show
snapshots of the simulation of growing nuclei seeded at a
bump, which display characteristic features in the free energy landscape for NG.  f) Free energy barriers for NG on a fixed geometry
and different degrees of supercoolings ($1/\epsilon$). $R_\textrm{q}$ represents a measure of the roughness of the substrate. For high
supercoolings ($1/\epsilon < 150$) the free energy display a
single barrier. The inset shows the
evolution of this barrier with the different supercoolings. The
presence of negative curvature induces the formation of local
minima an extra barrier for smaller supercoolings ($1/\epsilon
\sim 150-200$). Closer to coexistence ($1/\epsilon >250$) a new
minima and barrier appear as a consequence of the negative
curvature of the next nearest saddles. }
\end{figure*}
\end{center}

To unveil the dynamics of NG on more complex geometries, we
developed numerical simulations (see Methods) for geometries in which the surface height
at position $(x,y)$ is given by $Z=A \,cos(2\pi x/L) \, cos(2\pi
y/L)$ (see the the surface in Figs. 4a-e), where the
curvature is symmetrically distributed into positive and negative
curved regions.

Figure 4f shows the free energy landscape for nuclei initially
seeded on the top of a bump  (local positive curvature) for different
degrees of supercooling $1/\epsilon$. Here $\langle R \rangle$ acts as a collective coordinate and
is defined as the geodesic distance between the interface and the
center of the seed averaged along the perimeter, and $R_\textrm{q}$ is the root mean squared distributions of heights, being a parameter representing the substrate's roughness. Panels 4a-e show shape and sizes of nuclei inducing distinctive points in the free energy landscapes
for NG.

\begin{center}
\begin{figure}[t]
\includegraphics[width=7 cm]{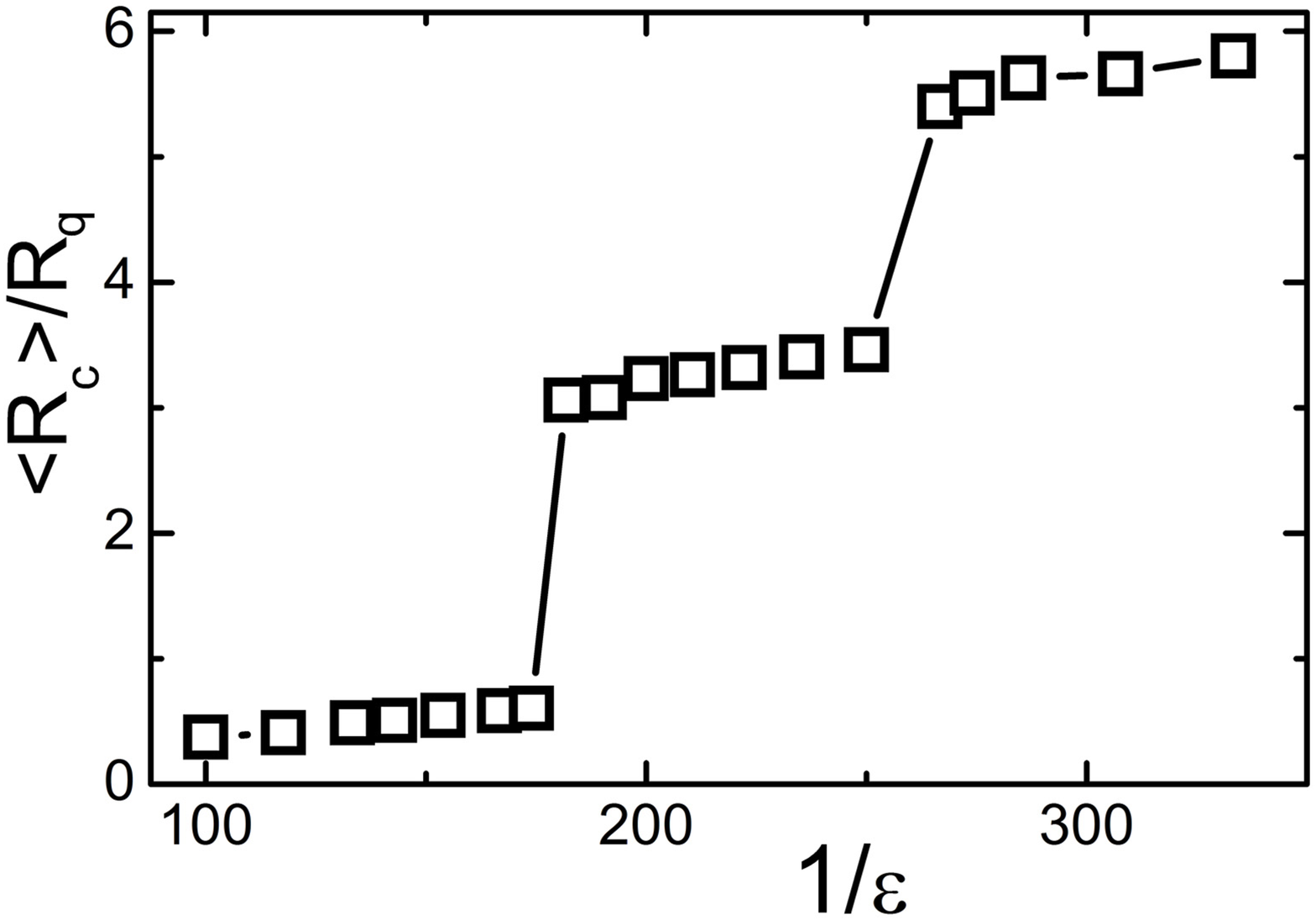}\caption{\textbf{Critical size for NG in a geometry of varying curvature.} As a consequence the varying
curvature in degree and sign, the critical size $R_\textrm{c}$ in sinusoidal geometries displays
discontinuous jumps when plotting as a function of the
supercooling $1/\epsilon$.}
\end{figure}
\end{center}

Figure 4f shows that for high supercoolings ($1/\epsilon \sim 100-150$)
the system displays a single barrier for NG, as found for
systems of constant curvature (plane, spheres, or pseudospheres).
The inset in Fig. 4f shows the expected enlargement of this barrier as the supercooling
decrease.

Remarkably, for low degrees of supercoolings multiple
barriers appear with local minima in between.
Starting from low supercooling a change of slope occurs around $\langle R \rangle/R_\textrm{q} \sim 2.5$
which gives a local minima for $1/\epsilon > 170$. This $\langle R \rangle$ corresponds to the
region in which the interface of the nuclei starts to approach the first neighboring
 saddles (see Fig. 4b for a nucleus of this size). Clearly, this is a
geometrical effect coming from an increase of surface energy of
nuclei due to the approach to a negative curved region.
Thus, initially propagating nuclei become
equilibrated and trapped by the negative curvature of the saddles.
The escape from these regions requires the overcoming of an extra
barrier associated with a much larger critical size $\langle R
\rangle/R_\textrm{q} \sim 3.5$ (Fig. 4c).

Here we note that for small supercooling on a periodically
curved substrate it is possible to obtain a microemulsion state as a metastable
thermodynamic phase.  These geometries are such that practically all nucleation events will
take place in regions of high positive curvature and the nuclei will be unable to invade regions
of negative curvature. Thus, at low supercooling the system will remain
arrested in a metastable microemulsion state which share
characteristics of both phases with
interesting perspective for applications. The formation of emulsion phases are typical of systems with a conserved order
parameter, like the familiar patterns obtained in oil-water
mixtures  \cite{Bray}. These are non-equilibrium states which evolve in time according to a self-similar
mechanism. However, on flat geometries emulsions can not be obtained for non-conserved order parameters, like
crystallization, because once a nucleus overcomes the critical size it freely
propagates throughout the system.

As the coexistence line is approached, the critical nuclei grow in
size, feeling the presence of other curved regions of the
substrate. For $1/\epsilon \sim 250-350$ a new local minimum
arises for nuclei of sizes around $\langle R \rangle /R_\textrm{q} \sim 5$.
This minimum corresponds to the trapping of nuclei by the negative
curvature of the next nearest saddle points (Fig. 4d), producing
the formation of an extra barrier associated to a larger
critical size $\langle R \rangle/R_\textrm{q} \sim 6$.

It is interesting to note here that this second geometry-induced barrier has a much
lower height as compared with the previous. This is just a consequence of how the curvature is
distributed on these sinusoidal substrates. For the second barrier
the negative curvature of the next nearest saddles is in part
compensated by the positive curvature of the nearest
positive-curved bumps, such that a lower activation is required to
escape from these regions. Thus, close to coexistence, the energy
landscape displays a "ratchet-like" form, where the system relaxes by
exploring subsequent metastable states of bigger size and smaller
activation energy.

Figure 5 shows the evolution of the
critical size $R_\textrm{c}$ for NG as a function of the degree of
supercooling ($1/\epsilon$). As expected, the critical size
decreases with supercooling. However, the formation of new minima
and extra barriers as supercooling is varied, originate abrupt
jumps in the critical size for nucleation.

Note that in complex geometries it is also possible that the
spatial location where the critical nuclei reside, could change as
supercooling is varied, originating thus not only jumps in
critical sizes, but also sudden changes in the surface locations where
critical nuclei develop.

\begin{center}
\begin{figure}[b]
\includegraphics[width=7.5 cm]{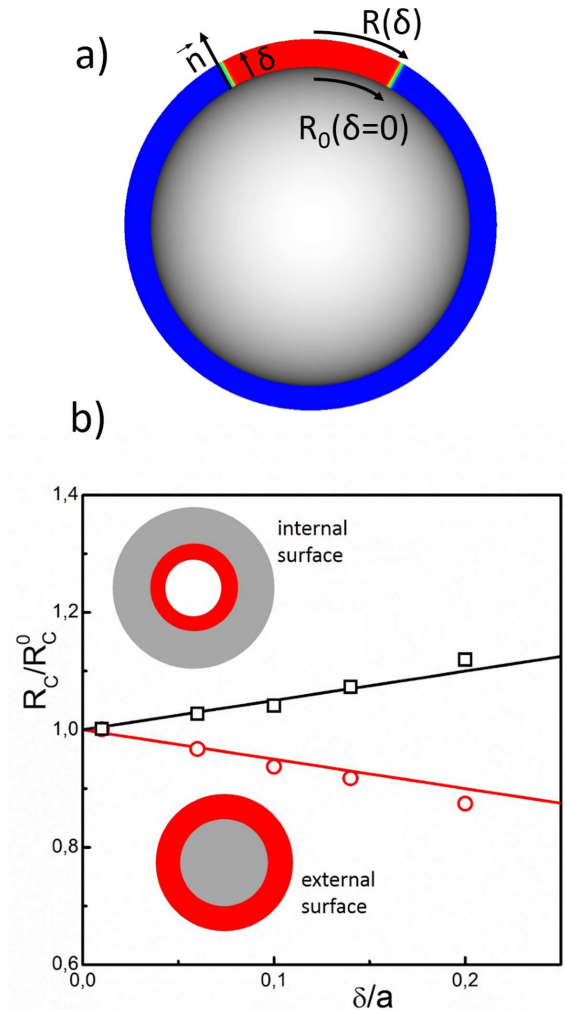}
\caption{\textbf{NG in 2D systems with thickness.} a) Cartoon showing an scheme of a nucleus in a surface with normal $\vec{n}$ and thickness $\delta$. Here while in the internal layer the nucleus has a radius $R_\textrm{0}\equiv R(\delta= 0)$, on the external layer the nucleus as a bigger size $R(\delta)>R_\textrm{0}$. b) Nucleation in spherical geometries of radius $a$ for systems with thickness $\delta$. The plot compares simulations (symbols) and theory (lines) for critical size $R_\textrm{c}$ (normalized with the critical size in a planar geometry $R_\textrm{c}^\textrm{0}$) as a function of thickness $\delta$. The data on the bottom (top) corresponds to NG on external (internal) surfaces, where $R_\textrm{c}$ decreases (grows) with $\delta$.}
\end{figure}
\end{center}

\textbf{Effect of the mean curvature.} Up to now we have considered the effects of the Gaussian curvature $K$ on NG. This is the only curvature which modifies the laws of NG in two-dimensional systems of negligible thickness. Here we consider physical systems where the thickness $\delta$ of nuclei can not be neglected, but where the NG process can still be considered as fundamentally 2D ($\delta<<R_\textrm{c}$). In such cases a system with thickness can be modelled by two parallel surfaces displaced by $\vec{\delta}=\delta \vec{n}$ \cite{SafranReview}, \cite{Hyde}, where $\vec{n}$ is the normal to the surface (see an scheme of the system in Figure 6a). As shown below, in such cases not only the Gaussian curvature $K$, but also the mean curvature $H$ plays an important role in NG.

The effects of finite thickness can be studied by considering the energy of formation of a nucleus:
\begin{equation}
\Delta F(r)=- V \Delta f+ A \, \sigma,
\end{equation}
where the volume $V$ and surface area $A$ can be calculated by considering the nucleus as composed of different layers. On each of these layers the section of the nucleus is (approximately) a geodesic circle, but due to the presence of curvature the radius of these circles will be in general a function of the thickness. Then, the volume and area can obtained by integrating the area and perimeter of the geodesic circles in the thickness of the system (see methods).

In the limit of small thickness ($\delta<<R$) and small curvatures ($R_\textrm{c}^\textrm{0}/R_\textrm{K}<<1$), the energy of formation of nuclei can be expanded leading to a simple expression for the critical size for NG (see methods):
\begin{equation}
R_\textrm{c}=R_\textrm{c}^\textrm{0} \,\Gamma_\textrm{KH}+\ldots,
\end{equation}
where again $\Gamma_\textrm{KH}$ represents a geometric factor, but now this factor is a function of both curvatures and thickness:
\begin{equation}
\Gamma_\textrm{KH}=1-\frac{1}{3} \, K_\textrm{0} \, {R_\textrm{c}^\textrm{0}}^2-\frac{1}{2}H_\textrm{0}\delta
\end{equation}
Thus, as previously found for the Gaussian curvature $K_\textrm{0}$, in general the presence of non-zero mean curvature $H_\textrm{0}$ and finite thickness $\delta$ will contribute to the formation of critical nuclei during NG.
Note however that as the mean curvature is related to the way the surface is embedded in 3D Euclidean space, the critical size for NG will be very different if the process develops on the external or internal surface of the substrate. For example, for the case of cylindrical substrates it will be much easier to nucleate in the external surface of the cylinder ($\Gamma_\textrm{KH}\equiv 1-\frac{1}{2}H_\textrm{0}\delta$) than on its internal face ($\Gamma_\textrm{KH}\equiv 1+\frac{1}{2} H_\textrm{0} \delta$). A similar behaviour was previously observed by using Monte Carlo simulations of NG dynamics in colloidal suspensions \cite{Frenkel2004}.

Figure 6b shows the good agreement between the theoretical expression Eqn. 15  and numerical simulations for NG on spheres for nuclei of different thickness $\delta$. Note how the critical size decrease with $\delta$ for NG on the external surface of the sphere, but increase with $\delta$ on an internal surface.

\begin{center}
\begin{figure}[t]
\includegraphics[width=8.5 cm]{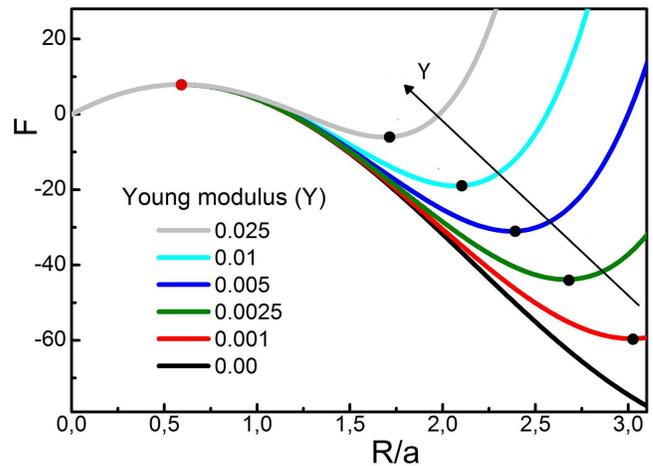}
\caption{\textbf{NG in a spherical substrate under elastic frustration.}  The plot shows the energy landscape $F$ for NG for different
Young modulus $Y$, where $R$ is the nuclei's geodesic radius and $a$ the sphere radius. For crystal patches having no defects the free energy display an absolute minima
related to the formation of nuclei reaching equilibrium sizes
(black dots) in addition to the conventional thermal barrier
(red dot).}
\end{figure}
\end{center}

\textbf{Effect of elastic frustration.} In general, when a crystal is forced to reside on a curved
surface some lattice bonds necessarily compress or stretch, giving
rise to an increase of the strain energy, geometrically frustrating the formation of crystal bonds \cite{Nelson}-\cite{VitelliPNAS}. In some
cases such elastic distortions can be released by introducing
topological defects in the lattice on particular regions of the
substrate.

In a recent experiment the effects of this elastic frustration on NG has been explored by following the dynamics of 2D
growing colloidal crystals on spherical substrates \cite{Meng}. Here the colloidal particles were constrained to reside on the curved interface of water droplets dispersed in oil. This study revealed the presence of an elastic instability, as a main consequence of the high geometric frustration in the crystallites. Due to the  rapid increase of the strain energy with size, the nuclei tend to grow in ramified fractal-like form on spheres, rather than in the typical compact structure observed in planar geometries.

The effects of curvature and frustration on crystallization has also been recently studied in experiments of 3D heterogeneous nucleation on the surface of curved substrates \cite{Gasser2014}, \cite{Sandomirski}. Here foreign particles are located inside colloidal suspensions, acting as seeds for crystallization, giving rise to a heterogeneous nucleation process which starts on the surface of a curved hard wall (the surface of the foreign particles) \cite{Lekkerkerker}. Such studies have also shown how curvature and geometric frustration can modify NG. It was found that although critical nuclei still form on the surface of the substrates, they tend to detach when growing, mainly because the curvature produce strains in the crystal lattices (some crystal's bonds would need to stretch or compress in order to adjust to the surface geometry), which largely increases the elastic free energy of growing nuclei. Although such NG processes are mainly 3D, and then in principle different to the 2D processes considered here, some similarities are found. For example due to the effects of mean curvature $H_\textrm{0}$ and thickness $\delta$ the NG is found to be different if the process develops on the outer or inner surface of the substrates.

The effects of elastic frustration can be incorporated in our model by adding the corresponding elastic free energy to the functional in Eq. 2. In general, the elastic energy of a curved crystalline monolayer on a curved
substrate can be written as:
\begin{equation}
F_\textrm{strain}=\frac{1}{2Y} \int d^2r \, \sqrt{g} (\Delta_\textrm{LB} \chi)^2,
\end{equation}
where $Y$ is the Young modulus of the crystal and $\chi$ is the Airy stress function which is dependent of both underlying curvature $K$ and presence and location of topological defects in the crystal.

In the case of spherical substrates of radius $a$, and for small nuclei of size $R$ containing no defects, the elastic free energy results in
$F_\textrm{strain}\approx \frac{Y \pi}{384} \frac{R^6}{a^4}$ \cite{Meng}. Figure 7 shows free energy barriers for NG on spherical substrates, for defect-free crystal nuclei of different
Young modulus. Here in addition to the thermal barrier (red dot) the fast increase in the strain of the nuclei with $R$ produces an absolute
minimum in the energy landscape (black dots), such that a growing nuclei would reach an equilibrium size, being unable to propagate throughout the system.
One way to escape from this energy minimum is by distorting the nuclei shape in less compact structures, as observed in the experiments of Meng et al. \cite{Meng}.

Note that for cases where the lattice structure is very
soft (where topological defects have low core energies), it is conceivable that the system would rapidly relax the
strain energy through the formation of topological defects. In
such cases, we can expect that the strain free energy contribution
do not play a major role in the nucleation dynamics and therefore
we can expect similar results for NG on curved geometries as
in previous sections.

\section*{Discussion}

In this work we have studied NG on curved geometries in systems where the
imbalance of volume and surface free energies of nuclei directs
the phase transition. It is precisely the change of this imbalance
with supercooling and local curvature that makes NG a rich and
complex phenomena in curved space.

In general, the critical size and propagation rate of nuclei are both strongly
dependent on the underlying geometry. For large curvatures the critical sizes can be severely modified (Fig. 3a). For example, for cases where the radius of curvature of the substrate is of the order of the critical size for nucleation on the plane ($ R_\textrm{c}^\textrm{0}/R_\textrm{K} \sim 1$), the critical size for nucleation is reduced in a $35\%$ for positive underlying curvature. Note however, that for the same thermodynamic conditions the critical size becomes about $250\%$ larger for negative underlying curvature.

Because of this effect, substrates of varying curvature impose complex
relaxation paths for NG, with several geometry-induced local minima, with interesting applications for phases deposited on substrates of controlled roughness. In general, the characteristic time $\tau_m$ that a system spends in a given metastable state roughly scales with its activation energy: $\tau_\textrm{m} \sim exp (\Delta F /k_\textrm{B}T)$
\cite{Kashchiev}, \cite{Kelton}. Contrary to NG of two-phase
systems in planar geometries, here the different local free energy
minima originated by the varying curvature would give rise to a
full spectrum of relaxation times. Such spectrum,
having the main information of how the relaxation process is
developed, is a fingerprint of the way the curvature is
distributed throughout the substrate.

Due to the generality and simplicity of our model, we expect
that the results obtained here could be applied to a variety of
systems in condensed matter, and other fields concerned with the
development of a new phase on a curved surface.
Experiments of colloidal crystallization on curved surfaces could be carried out on capillary bridges of negative Gaussian curvature and null mean curvature, used in the studies of 2D curved crystalline structures and defects in Refs. \cite{IrvineNature}, \cite{IrvineNatMat}.  The combined effects of Gaussian $K$ and mean $H$ curvatures and thickness $\delta$ on NG could also be experimentally addressed by confining block copolymers, liquid crystals or similar systems into different geometries \cite{Shi}, like corrugated substrates \cite{HexemerThesis}, \cite{VegaSoftmatter}, spherical shells \cite{FernandezNievesNatPhys} or toroidal droplets \cite{FernandezNievesPNAS}. In all these systems we expect that the interplay between geometry and phase nucleation will lead novel pattern formation phenomena that can be controlled to design useful supramolecular and soft materials.

\subsection*{Acknowledgements}

This work was supported by the National Research Council of
Argentina, CONICET, ANPCyT, and Universidad Nacional del Sur.

\subsection*{Author contributions}

LRG and DAV conceived the work. LRG developed the model and carried out analytical
calculations. NAG, LRG and DAV performed the simulations. All authors discussed and
analized the results. LRG, VV, JL and DAV contributed to the manuscript preparation.

\subsection*{Competing financial interests}

The authors declare no competing financial interests.

\section*{Methods}

\textbf{Equations describing the steady propagation of nuclei.}
After the functional derivation in Eqn. 4, the evolution equation for
a non-conserved order parameter takes the form:
\begin{equation}
\label{evolutionequation} \frac{\partial \psi}{\partial
t}=D\Delta_\textrm{LB}\psi-\frac{\partial f(\psi)}{\partial \psi}
\end{equation}
Upon using geodesic polar coordinates, the Laplace-Beltrami
operator can be written as \cite{faraudo}:
\begin{equation}
\Delta_\textrm{LB} \psi=\frac{\partial^2
\psi}{\partial r^2}+\kappa_\textrm{g}(r,\theta)\frac{\partial
\psi}{\partial r}+\frac{1}{r^2}\frac{\partial^2 \psi}{\partial
\theta^2}\exp[2 \int_0^r \kappa_\textrm{g}(u)du],
\end{equation}
where $\kappa_\textrm{g}(r,\theta)=\frac{\partial }{\partial r}
ln\sqrt{G(r,\theta)}$ is the local curvature of the geodesic
circles. For general geometries the Laplace-Beltrami operator
takes a complex form, being dependent on both coordinates
$(r,\theta)$. However, for surfaces with radial symmetry respect
to the origin $O$, the geodesic curvature is only a function of
the geodesic distance $\kappa_\textrm{g}\equiv\kappa_\textrm{g}(r)$. In such cases
we expect the nuclei to have radial symmetry, such that the
Laplace-Beltrami operator takes the much simplified form $\Delta_\textrm{LB} \psi \equiv
\frac{\partial^2 \psi}{\partial
r^2}+\kappa_\textrm{g}(r,\theta)\frac{\partial \psi}{\partial r}.$

We can now describe steady state solutions representing the growth or
collapse of nuclei during NG. Here the motion of a steadily propagating
interface can be described through a single combined coordinate
$X=r-R(t)$, where $R(t)$ is the radius of the nuclei
at time $t$. We assume that the interface is narrow respect to the nuclei radius $R$. Then we can take $\kappa_\textrm{g}(r) \equiv \kappa_\textrm{g}(R)$
and rewrite the evolution equation in the comoving coordinates $X$, as the system of coupled equation Eqns. 6 and 7 in the results section.

\textbf{Simulations.} The simulations of NG in curved space performed in this work were
developed by seeding an initial (geodesic) circular nucleus of a
prescribed initial radius $R_\textrm{0}$ on a particular region of the substrate,
and studying the temporal evolution through the numerical
resolution of Eqn. \ref{evolutionequation}.

The initial circular nuclei were obtained, for any underlying geometry, through a fast marching
algorithm \cite{Sethian}. Equation \ref{evolutionequation} can be
accurately solved through a finite difference scheme, forward in
time and centered in space and periodic boundary conditions \cite{GomezSmectics},\cite{Garcia}. In the case of systems with finite thickness the simulations were performed by numerically solving Eqn. \ref{evolutionequation} through finite elements, with periodic and null flux boundary conditions.

For different geometries and degrees of supercoolings $\varepsilon$ the critical sizes $R_c$ were obtained by finding the smallest nuclei which is able to grow in time. The free energy landscapes for NG were obtained by inserting the order parameter distribution $\psi(\textrm{\textbf{r}})$, of growing or collapsing nuclei of size $R(t)$, in the free energy functional Eqn. 2.

\textbf{Critical size in systems with thickness.} Given a nucleus in a curved system with thickness, some general conclusions can be drawn in the limit of small width $\delta$ and small critical size $R_\textrm{c}$.

To get the volume $V$ and area $A$ associated to the formation during nucleation (Eqn. 14), the nucleus is considered as having different layers, where in each layer the nucleus has a  circular shape. Thus, the volume and area can be obtained by integrating the area and perimeter of the geodesic circles in the thickness of the system.  Here the perimeter $P_{\delta}$ and area $A_{\delta}$ of these circles can be written in the approximate form \cite{Struik}:
\begin{eqnarray}
P_{\delta}&\approx& 2 \pi R(\delta)-\frac{\pi}{3} K(\delta) R(\delta)^{3}\\
A_{\delta}&=& \pi R(\delta)^{2}-\frac{\pi}{12} K(\delta) R(\delta)^{4}
\end{eqnarray}
where $R(\delta)$ and $K(\delta)$ are the radius and Gaussian curvature in the $\delta$-layer, given by:
\begin{eqnarray}
R(\delta)&\sim& (1+H_\textrm{0} \delta) R_\textrm{0}\\
K(\delta)&=& \frac{K_\textrm{0}}{1+2H_\textrm{0} \delta+K_\textrm{0} \delta^2}\sim\frac{K_\textrm{0}}{1+2H_\textrm{0}\delta}
\end{eqnarray}
Here $R_\textrm{0}$, $K_\textrm{0}$ and $H_\textrm{0}$ represent the size of nuclei and the Gaussian and mean curvatures evaluated on the surface of the substrate (for $\delta\equiv0$). The above expression for the Gaussian curvature is a well-known result which relates the Gaussian curvature of a surface $K$ with the Gaussian $K_\textrm{0}$ and mean $H_\textrm{0}$ curvatures of a parallel surface displaced by $\vec{\delta}=\delta \vec{n}$, with $\vec{n}$ the normal to the surface \cite{SafranReview}, \cite{Hyde}.

In this approximation the area and volume of the nucleus take the simple forms:
\begin{eqnarray}
A&\approx& (2 \pi R_\textrm{0}-\frac{\pi}{3} K_\textrm{0} R_\textrm{0}^{3})(1+\frac{1}{2}H_\textrm{0}\delta)\delta\\
V&\approx& (\pi R_\textrm{0}^{2}-\frac{\pi}{12} K_\textrm{0} R_\textrm{0}^{4})(1+H_\textrm{0}\delta)\delta
\end{eqnarray}
The replacement of these expressions in the energy of formation of a nucleus with thickness (Eqn. 14), straightforwardly leads to the expression of the critical size for NG (Eqn. 15).

\end{document}